\pdfoutput=1
\documentclass[10pt, a4paper]{article}
\usepackage[left=2cm, right=2cm, top=3cm, bottom=3cm]{geometry}
\usepackage[utf8]{inputenc}
\usepackage[english]{babel}

\usepackage{xcolor}
\definecolor{bluer}{RGB}{50, 50, 150}
\definecolor{redr}{RGB}{220, 50, 50}
\definecolor{pinkr}{RGB}{200, 0, 100}

\usepackage{amsmath}
\usepackage{amsfonts}
\usepackage{braket}
\usepackage{mathtools}
\usepackage{lmodern}
\usepackage[
    colorlinks=true,
    citecolor=bluer,
    linkcolor=redr,
    urlcolor=pinkr,
]{hyperref}
\usepackage{parskip}

\usepackage{multicol}
\usepackage{stmaryrd}
\usepackage{mathrsfs}
\usepackage{dsfont}
\usepackage{tikz}
\usepackage{xurl}
\usepackage{changepage}
\usepackage[symbol,hang,flushmargin]{footmisc}
\usepackage[explicit]{titlesec}
\usepackage{caption}
\usetikzlibrary{decorations.pathmorphing}

\titleformat{\section}{\normalsize\bfseries}{\thesection.}{0.4em}{\MakeUppercase{#1}}
\titleformat{\subsection}{\normalsize\bfseries}{\thesubsection.}{0.4em}{#1}

\usepackage[backend=bibtex, style=numeric-comp, sorting=none, maxbibnames=99, hyperref=true, url=false, doi=false]{biblatex}
\newbibmacro{string+url}[1]{\href{\thefield{url}}{#1}}
\DeclareFieldFormat*{title}{\usebibmacro{string+url}{#1}}
\DeclareFieldFormat*{journaltitle}{\mkbibitalic{#1}\addcomma}
\renewbibmacro{in:}{}
\AtEveryBibitem{%
  \clearfield{month}%
  \clearfield{pages}%
}

\DeclareFieldFormat[book,inproceedings,unpublished]{date}{(#1)}
\DeclareListFormat[book,inproceedings]{publisher}{\mkbibitalic{#1}\nopunct}
\DeclareFieldFormat[inproceedings]{booktitle}{\mkbibitalic{#1}\addcomma}
\DeclareFieldFormat[misc]{url}{\href{#1}{#1}\nopunct}
\DeclareFieldFormat[unpublished]{note}{\mkbibitalic{#1}}

\DeclareBibliographyDriver{misc}{%
    \usebibmacro{begentry}%
        \printfield{url}%
    \usebibmacro{finentry}
}

\DeclareBibliographyDriver{unpublished}{%
    \usebibmacro{begentry}%
        \printnames{author}\adddot\space%
        \printfield{title}\adddot\space%
        \printfield{note}\space%
        \printdate\adddot%
    \usebibmacro{finentry}
}

\usepackage{xcolor}
\definecolor{darkred}{RGB}{220, 0, 0}

\usepackage{xspace}
\newcommand{\dd}{\ensuremath{\mathrm{d}}}
\DeclarePairedDelimiterXPP\tmpTr[1]{\mathrm{Tr}}{[}{]}{}{#1}
\newcommand{\Tr}{\tmpTr*}
\DeclarePairedDelimiterXPP\tmpE[1]{\mathbb{E}}{[}{]}{}{#1}
\newcommand{\E}{\tmpE*}
\DeclarePairedDelimiterXPP\tmppr[1]{\mathbb{P}}{[}{]}{}{#1}
\newcommand{\pr}{\tmppr*}
\newcommand{\dt}{\ensuremath{\dd t}}
\newcommand{\dYt}{\ensuremath{\dd Y_t}}
\newcommand{\dWt}{\ensuremath{\dd W_t}}
\newcommand{\dNt}{\ensuremath{\dd N_t}}
\newcommand{\rhobar}{\ensuremath{\overline{\rho}}}
\newcommand{\Id}{\ensuremath{\mathrm{I}}}
\newcommand{\given}{\ensuremath{\:\vert\:}}
\newcommand{\Ld}{\ensuremath{L^\dag}}
\newcommand{\Lcal}{\mathcal{L}}
\newcommand{\Tcal}{\mathcal{T}}
\newcommand{\Zcal}{\mathcal{Z}}
\newcommand{\Ls}{\mathscr{L}}
\newcommand{\Lp}{L_+}
\newcommand{\Lt}{L_\times}
\newcommand{\dodt}{\frac{\dd}{\dt}}
\newcommand{\1}{\mathds{1}}
\newcommand{\pa}[1]{\partial_{\alpha_{#1}}}

\begin{document}

\begin{center}
    {\fontsize{14}{0}\textbf{
        Correlation functions for realistic continuous quantum measurement
    }}
    \\\vspace{0.5cm}
    {\large
        Pierre Guilmin\textsuperscript{1,2,}\footnote[1]{\href{mailto:pierre.guilmin@alice-bob.com}{pierre.guilmin@alice-bob.com}}, Pierre Rouchon\textsuperscript{2} and Antoine Tilloy\textsuperscript{2}
    }
    \\\vspace{0.2cm}
    {\normalsize
        \textsuperscript{1}\textit{Alice \& Bob, 53 Bd du Général Martial Valin, 75015 Paris, France}
        \\
        \textsuperscript{2}\textit{Laboratoire de Physique de l’École Normale Supérieure, Mines Paris,}
        \\
        \textit{Inria, ENS, Université PSL, Sorbonne Université, Paris, France}
        \\\vspace{0.2cm}
        May 22, 2023\,\footnote[2]{\copyright\ 2023 the authors. This work has been accepted to IFAC for publication under a Creative Commons Licence CC-BY-NC-ND.}
        \\\vspace{0.5cm}
    }
\end{center}

\begin{adjustwidth}{2cm}{2cm}
    We propose a self-contained and accessible derivation of an exact formula for the $n$-point correlation functions of the signal measured when continuously observing a quantum system. The expression depends on the initial quantum state and on the Stochastic Master Equation (SME) governing the dynamics. This derivation applies to both jump and diffusive evolutions and takes into account common imperfections of realistic measurement devices. We show how these correlations can be efficiently computed numerically for commonly filtered and integrated signals available in practice.\\
\end{adjustwidth}

\begin{multicols}{2}

\section{Introduction}
Experiments studying quantum systems generally follow the \emph{prepare, evolve and measure} pattern: the system is first prepared in a known quantum state, then it evolves unobserved for a certain period of time, and finally a projective measurement is performed. However, it was discovered in the 1990s that certain experimental setups allow to continuously measure a quantum system while it evolves \cite{wiseman2009quantum}. Nowadays, continuous measurements are frequently used by experimenters, notably in superconducting quantum circuits \cite{weber2016quantum,ficheux2018dynamics}. In these experiments, the observer is constantly acquiring information about the state of the system, and the impact of the measurement back-action must be taken into account at every time. The dynamics of such a continuously measured quantum system is described by the \emph{Stochastic Master Equation} (SME) formalism.

The quantum state and/or parameters can be reconstructed from the measured signal via quantum filtering \cite{wiseman2009quantum}, but this is usually computationally too expensive. In practice, the experimenters have direct access to the measured signal, so they can trivially calculate its $n$-point correlation functions. We show in this paper how these functions can be expressed explicitly from the SME modelling the system. Thus, albeit not optimal from a Bayesian point of view, they are an alternative and more practical approach than quantum filtering for quantum state reconstruction or for parameter estimation (see \textit{e.g.} \cite{campagne2016observing,six2015parameter}).

Several recent and older works calculate these functions analytically in restricted cases \cite{barchielli1991measurements,korotkov2001output,wiseman2009quantum,xu2015correlation,diosi2016structural,jordan2016anatomy,foroozani2016correlations,atalaya2018multitime}, and a general derivation in the case of the diffusive SME was discovered independently by \cite{hagele2018higher} and \cite{tilloy2018exact} in 2018. Related calculations can also be found in the field of condensed matter physics, which focuses on full counting statistics, cumulants and spectral representation \cite{flindt2010counting,sifft2021quantum,landi2023current}. This paper combines the general method developed in \cite{tilloy2018exact} to derive $n$-point correlations for diffusive SME with their discrete-time formulations presented in \cite{rouchon2022tutorial}. It provides explicit formulae of $n$-point correlations for both jump and diffusive SME: formulae~\eqref{eq:jump-2pt} and~\eqref{eq:diff-2pt} for 2-point; \eqref{eq:jump-npt} and~\eqref{eq:diff-npt} for $n$-point. Computing these formulae for realistic signals involve solutions of modified Lindblad master equations as shown in equations~\eqref{eq:rho0} to~\eqref{eq:rho12}. As far as we know, such calculations are not available in the literature, in particular for the jump SME. Furthermore, we detail all the ingredients necessary for the practical calculation of these $n$-point correlation functions on arbitrary quantum systems, including detector imperfections and for realistic (binned or filtered) experimental data. For clarity, the presentation focuses mainly on a single detector, but we also give the formula to derive $n$-point correlations between diffusive and/or jump signals coming from multiple different detectors.

This paper is organised as follows. In section~\ref{sec:2}, we recall the structure of continuous-time jump or diffusive SMEs, and we present their discrete-time formulations based on partial Kraus maps. In section~\ref{sec:3}, we derive the general correlations formula from the underlying SME and classical post-filtering. This derivation is almost straightforward in the discrete-time formulation, and it directly provides explicit formulae in the continuous-time formulation. In section~\ref{sec:4} we detail a novel numerical method for practically computing the correlation functions. Section~\ref{sec:5} briefly introduces an example application.

\section{The SME formalism} \label{sec:2}
A SME is a \emph{non-linear and non-determinsitic} differential equation which determines the evolution of the system state $\rho_t$ at time $t$, conditioned on the detector measurement record. Different measurement schemes lead to different types of evolution: the state can evolve discontinuously, with sudden jumps at random time, which is modeled by a \emph{jump} SME, or continuously in state space, which is modeled by a \emph{diffusive} SME. In quantum optics for instance, the first situation corresponds to photon counting schemes \cite{gardiner2004quantum} and the second to homodyne or heterodyne detection schemes \cite{wiseman1993quantum}.

In this section, we first introduce the jump and diffusive SMEs, then we give an equivalent discrete-time formulation, and we conclude by explaining the link between SMEs and the \emph{linear and deterministic} Lindblad master equation describing unobserved quantum systems.

\subsection{Jump SME}
When the result of the measurement at time $t$ is either a detection event or a no-detection event, the system state undergoes abrupt jumps from one state to another upon detection. The \emph{discrete-valued continuous-time} stochastic process driving the SME is the increment $\dNt$, taking the value $0$ for no-detection and $1$ for detection with probabilities depending on the system state at time $t$:
\begin{align}
    \pr{\dNt=0} &= 1 - \pr{\dNt=1}, \\
    \pr{\dNt=1} &= \left(\theta + \eta\Tr{L\rho_t\Ld}\right)\dt,
\end{align}
where $L$ is an arbitrary operator characterising the detector, $\theta\geq0$ is the dark count rate and $0<\eta\leq1$ is the detector efficiency.

The evolution of $\rho_t$ is described by the jump SME \cite{rouchon2022tutorial}:
\begin{align}\label{eq:jump-sme}
    \begin{split}
    \dd\rho_t =& -i[H,\rho_t]\dt + \mathcal{D}[L](\rho_t)\dt \\
               &+ \mathcal{G}[L](\rho_t)\left(\dNt - \left(\theta + \eta\Tr{L\rho_t\Ld}\right)\dt\right),
    \end{split}
\end{align}
where $H$ is the system Hamiltonian, $\mathcal{D}[L](\rho) = L\rho\Ld - \frac{1}{2}\Ld L\rho - \frac{1}{2}\rho\Ld L$ is the standard dissipator, and the superoperator $\mathcal{G}[L]$ describing the back-action of the measurement is defined by:
\begin{equation}
    \mathcal{G}[L](\rho) = \frac{\theta\rho + \eta L\rho\Ld}{\theta + \eta\Tr{L\rho\Ld}} - \rho.
\end{equation}

The \emph{continuous-time signal} measured by the detector is $I_t=\dYt/\dt$, where $\dYt$ is directly defined as the stochastic increment in the case of the jump SME:
\begin{equation}\label{eq:jump-dsignal}
    \dYt = \dNt.
\end{equation}
The signal $I_t$ is the rate of change of the counting process $N_t$, which counts the number of jumps that occurred in the time interval $[0,t]$.

An experiment corresponds to a specific realisation of the stochastic process $\dNt$, giving rise to a \emph{quantum trajectory} which describes the path followed by the state of the quantum system over time. This trajectory is conditioned on the measurement results: we can replace the stochastic term $\dNt$ by the measured signal values at each time in~(\ref{eq:jump-sme}), and thus reconstruct $\rho_t$ from the measurement record.

\subsection{Diffusive SME}
When the result of the measurement at time $t$ takes a continuous range of values, the system state evolves continuously in state space. The \emph{real-valued continuous-time} stochastic process driving the SME is the Wiener process $\dWt$, taking independent Gaussian distributed increment.

The evolution of $\rho_t$ is described by the diffusive SME in Itô form \cite{jacobs2006straightforward}:
\begin{align}\label{eq:diff-sme}
    \begin{split}
        \dd\rho_t =&~ -i[H,\rho_t]\dt + \mathcal{D}[L](\rho_t)\dt\\
        &+ \sqrt{\eta}\mathcal{M}[L](\rho_t)\dWt,
    \end{split}
\end{align}
where the superoperator $\mathcal{M}[L]$ describing the back-action of the measurement is defined by:
\begin{equation}
    \mathcal{M}[L](\rho) = L\rho + \rho\Ld - \Tr{(L+\Ld)\rho}\rho.
\end{equation}

Similarly, the continous-time signal measured by the detector $I_t=\dYt/\dt$ verifies:
\begin{equation}\label{eq:diff-dsignal}
    \dYt = \sqrt{\eta}\Tr{(L+\Ld)\rho_t}\dt + \dWt.
\end{equation}
Sometimes the signal is defined with a different but equivalent normalisation $\dYt'=\dYt/(2\sqrt\eta)$.

As for the jump SME, an experiment corresponds to a specific realisation of the stochastic process $\dWt$, giving rise to a quantum trajectory that can be reconstructed from the measured signal.

\subsection{Discrete-time formulation}
We can derive both SMEs by taking the limit of infinitely frequent and infinitely weak projective measurements \cite{attal2006repeated,attal2010stochastic}. We use such a discrete-time picture with a slightly different formulation as described in \cite{rouchon2022tutorial}.

In this formulation, the measurement process is described by a \emph{quantum instrument}, which combines a quantum measurement characterised by a \emph{positive operator-valued measure} (POVM) and a classical uncertainty on the measurement result accounting for imperfections of the detector. A map depending on the measurement result is applied at each small time step $\dt$:
\begin{equation}\label{eq:povm}
    \rho_{k+1} = \frac{K_{r_{k+1}}(\rho_k)}{\Tr{K_{r_{k+1}}(\rho_k)}},
\end{equation}
where $\rho_k$ is the state of the system at step $k$ (at time $t=k\dt$), and $K_{r_{k+1}}$ is a linear map depending on $r_{k+1}$, the measurement result at step $k+1$.

For both SMEs, we define the corresponding quantum instrument by specifying the linear map $K_r$ associated with each possible measurement result $r$. The continuous-time formulations~(\ref{eq:jump-sme}) and~(\ref{eq:diff-sme}) are recovered by taking the limit of infinitesimally small time step, and expanding the expression $\rho_{k+1} - \rho_k \sim \rho_{t+\dt} - \rho_t = \dd\rho_t$ to first order in $\dt$.

\textbf{Jump SME} -- For the jump SME, the measurement result is either $0$ or $1$, and the corresponding maps are:
\begin{align}
    K_0(\rho) &= (1-\theta\dt)M_0\rho M_0^\dag + (1-\eta)M_1\rho M_1^\dag,\\
    K_1(\rho) &= \theta\dt M_0\rho M_0^\dag + \eta M_1\rho M_1^\dag,
\shortintertext{with}
    M_0 &= \Id - iH \dt - \frac{1}{2}\Ld L \dt, \\
    M_1 &= L\sqrt{\dt}.
\end{align}

The probability of obtaining the measurement result $r$ at step $k+1$ depends only on the preceding state $\rho_k$:
\begin{equation}\label{eq:povm-discrete-proba}
    \pr{r_{k+1}=r \given \rho_k} = \Tr{K_r(\rho_k)}.
\end{equation}

We can evaluate the probability of obtaining the measurement record $\{r_1,r_2\}$ knowing the initial state $\rho_0$:
\begin{equation}
    \pr{r_1, r_2 \given \rho_0} = \pr{r_1 \given \rho_0} \times \pr{r_2 \given r_1, \rho_0}\notag.
\end{equation}
The first term is given by~(\ref{eq:povm-discrete-proba}): $\pr{r_1 \given \rho_0} = \Tr{K_{r_1}(\rho_0)}$. To evaluate the second term we use~(\ref{eq:povm}) to write the state at step $k=1$ conditioned on the measurement result $r_1$: $\rho_1 = K_{r_1}(\rho_0)/\Tr{K_{r_1}(\rho_0)}$. Then using~(\ref{eq:povm-discrete-proba}) again:
\begin{equation}
    \pr{r_2 \given r_1, \rho_0} = \Tr{K_{r_2}(\rho_1)} = \Tr{K_{r_2} \left(\frac{K_{r_1}(\rho_0)}{\Tr{K_{r_1}(\rho_0)}}\right)}\notag.
\end{equation}
Combining the two terms, we get:
\begin{equation}
    \pr{r_1, r_2 \given \rho_0} = \Tr{K_{r_2}K_{r_1}(\rho_0)}.
\end{equation}

This result directly extends to the probability of obtaining the measurement record $\{r_1, \dots, r_N\}$:
\begin{equation}\label{eq:jump-povm-record-proba}
    \pr{r_1, \dots, r_N \given \rho_0} = \Tr{K_{r_N}\dots K_{r_1}(\rho_0)}.
\end{equation}

\textbf{Diffusive SME} -- For the diffusive SME, the measurement result takes a continuous range of values and the map corresponding to the measurement result $r$ is:
\begin{align}\label{eq:diff-povm-K}
    K_r(\rho) &= M_r \rho M_r^\dag + (1-\eta)L\rho\Ld\dt,
\shortintertext{with}
    \label{eq:diff-povm-kraus}
    M_r &= \Id - iH \dt - \frac{1}{2} \Ld L \dt + \sqrt{\eta}L r.
\end{align}

The equivalent of~(\ref{eq:povm-discrete-proba}) is given by the probability density to get a measurement result in $[r,r+\dd r[$ at step $k+1$ knowing the state $\rho_k$:
\begin{equation}
    \dd\mathbb{P}\big[r_{k+1} \in [r, r+\dd r[ \given \rho_k\big] = \dd \mu(r) \Tr{K_r(\rho_k)},
\end{equation}
where $\dd\mu(r)$ is the Gaussian measure centered on $0$ with variance $\dt$ \cite{jacobs2006straightforward,rouchon2022tutorial}:
\begin{equation}\label{eq:gaussian-measure}
    \dd\mu(r) = \frac{1}{\sqrt{2\pi\dt}}\exp\left(\frac{-r^2}{2\dt}\right) \dd r.
\end{equation}

The same calculations as for the jump SME give the probability density to get the measurement record $\{r_1,\dots,r_N\}$:
\begin{align}\label{eq:diff-povm-record-proba}
    \begin{split}
        &\dd\pr{r_1, \dots, r_N \given \rho_0} =\\
        &\qquad \dd\mu(r_1) \dots \dd\mu(r_N) \Tr{K_{r_N} \dots K_{r_1}(\rho_0)}.
    \end{split}
\end{align}

\subsection{Unconditioned evolution}
When the measurement results are unknown to the observer, for example for a purely dissipative process or for unread measurements, the system dynamics is deterministic. The evolution of the \emph{unconditioned state} $\rhobar_t$ is recovered by averaging over all possible quantum trajectories --- or equivalently over all possible measurement records --- weighted by their probability of occurrence:
\begin{equation}
    \rhobar_t = \E{\rho_t},
\end{equation}
where $\mathbb{E}$ denotes the statistical average over the stochastic process driving the SME. Note that the unconditioned state does not depend on the stochastic process averaged over (jump or diffusive): different types of stochastic evolution lead to the same ensemble average trajectory.

The evolution of $\rhobar_t$ is then described by the \emph{linear and deterministic Lindblad master equation} \cite{haroche2006exploring}:
\begin{equation}
    \frac{\dd\rhobar_t}{\dt} = -i[H,\rhobar_t] + \mathcal{D}[L](\rhobar_t) = \Lcal(\rhobar_t),
\end{equation}
where $\Lcal$ is the system Lindbladian, the superoperator generating the evolution of the system when the observer does not know the measured signal. For time-independent Lindbladian, the formal solution reads:
\begin{equation}
    \rhobar_t = e^{t\Lcal}(\rho_0).
\end{equation}
For time-dependent Lindbladian $\Lcal_t$, the solution is written using the time-ordered exponential:
\begin{equation}
    \rhobar_t = \Tcal \exp\left(\int_{0}^{t} \Lcal_{t'}\dd t'\right) (\rho_0),
\end{equation}
where $\Tcal$ is the time-ordering symbol.

Similarly to the jump and diffusive SME, the Lindblad master equation has a discrete-time formulation. The general evolution of an \emph{unobserved open quantum system} between two time steps is characterised by a \textit{completely positive trace preserving} (CPTP) linear map $K$ (also called \emph{quantum channel} or \emph{dynamical map}):
\begin{equation}
    \rhobar_{k+1} = K(\rhobar_k).
\end{equation}

In this discrete-time formulation, we also recover the unconditionned evolution by averaging over all possible measurement outcomes at each step:
\begin{equation}
    \rhobar_{k+1} = \E{\rho_{k+1} \given \overline{\rho}_k}.
\end{equation}

As in the continuum, we find the same CPTP map $K$, that is the same unconditionned state dynamics, when averaging over either of the stochastic processes (jump or diffusive):
\begin{equation}
    K(\rho) = K_0(\rho) + K_1(\rho) = \int_{-\infty}^{\infty} \dd\mu(r) K_{r}(\rho).
\end{equation}
In the continuous-time limit we recover the evolution generated by the system Lindbladian $\Lcal$.

\section{Correlation functions}\label{sec:3}
The statistics of the measured signal are fully characterised by its correlation functions. The $n$-point correlation function of the signal $I_t$ is defined by:
\begin{equation}
    C_{t_1, t_2, \dots, t_n} = \E{I_{t_1} I_{t_2} \dots I_{t_n} \given \rho_0}.
\end{equation}
The one-point correlation function is the signal mean, and the two-point correlation function is the signal autocorrelation.

The signal $I_t$ is a singular quantity, in the case of the jump SME it can be loosely thought of as a series of Dirac delta distributions at the times of detection, and in the case of the diffusive SME as white noise with a trend. This quantity is better defined when it is integrated against a \emph{smooth test function} $f$:
\begin{equation}
    I_f = \int f_t \dYt.
\end{equation}
In practice, the signal is obtained from a finite bandwidth detection chain, and is therefore effectively filtered. Experimentally, the smooth test function $f$ then corresponds to the transfer function of the detection chain. In the following, we will refer to $I_t$ as the \emph{sharp} signal, and to $I_f$ as the \emph{filtered} signal.

The filtered signal $I_f$ is the only quantity actually available to an experimenter. Thus, we are also interested in calculating its correlation functions:
\begin{equation}
    C_{f_1, f_2, \dots, f_n} = \E{I_{f_1} I_{f_2} \dots I_{f_n} \given \rho_0}.
\end{equation}

The objective of this section is to give an analytical formula depending only on the SME for the correlation functions of the sharp and filtered signal. We first introduce the generating functional of the correlation functions, allowing us to evaluate both $C_{t_1, \dots, t_n}$ and $C_{f_1, \dots, f_n}$. We then deduce the formula for the correlation functions using the discrete-time formulation and its continuous limit. The remainder of the section is devoted to examples of how to calculate basic correlation functions using this general formula, and we finally generalise the result to the case of multiple detectors and mixed jump-diffusive SME.

\subsection{Generating functional}
The sharp and filtered signal correlation functions can both be determined using the \emph{generating functional} $\Zcal(j)$ defined by \cite{barchielli2009quantum,tilloy2018exact}:
\begin{equation}
    \Zcal(j) = \E{\exp\left(\int_{u=0}^{u=T} j_u\dd Y_u\right) \Big\vert\: \rho_0},
\end{equation}
where $j$ is a smooth test function and $T$ is a large time, typically larger than any time involved in the correlation functions we wish to evaluate. This generating functional is defined analogously to the moment-generating function $M(t)=\E{e^{tX}}$ of a random variable $X$, whose $n$-th derivative yields the $n$-th moment of $X$:
\begin{equation}
    \E{X^n}=\frac{\dd^n}{\dt^n}M(t)\Big\vert_{t=0}.
\end{equation}

\textbf{Sharp signal correlations} -- The functional derivative of $\Zcal(j)$ with respect to $j_t$ for $t\in[0,T]$ reads:
\begin{equation}
    \frac{\delta}{\delta j_t} \Zcal(j) = \E{\frac{\dYt}{\dt}\exp\left(\int_0^T j_u\dd Y_u\right) \Big\vert\: \rho_0}.
\end{equation}

Thus for $t_1 < t_2 < \dots < t_n$, the sharp signal correlation function can be expressed as:
\begin{equation}
    C_{t_1, \dots,t_n} = \frac{\delta}{\delta j_{t_1}} \cdots \frac{\delta}{\delta j_{t_n}} \Zcal(j) \Big\vert_{j=0}.
\end{equation}

The correlation functions involving \emph{equal time contributions} are not well defined for the sharp signal, because they yield Dirac delta distribution. However, these contributions should not be missed when evaluating the filtered signal correlation functions.

\textbf{Filtered signal correlations} -- The correlation functions of the filtered signals $I_{f_1},\dots,I_{f_n}$ are given by the standard partial derivative of $\Zcal(\alpha_1 f_1 + \dots + \alpha_n f_n)$ with respect to $\alpha_1,\dots,\alpha_n$:
\begin{equation}
    C_{f_1, \dots, f_n} = \frac{\partial}{\partial \alpha_1} \cdots \frac{\partial}{\partial \alpha_n} \Zcal(\alpha_1 f_1 + \dots + \alpha_n f_n) \Big\vert_{\alpha_1,\dots,\alpha_n=0}.\label{eq:filtered-correlation}
\end{equation}

\subsection{Derivation of the analytical formula}
Our goal is to find an analytical formula for the generating functional $\Zcal(j)$. In this subsection we give the proof for the jump SME by using the discrete-time formulation, and explain how it extends to the diffusive SME. For a calculation in the diffusive case relying only on continuous stochastic calculus techniques without resorting to discretisation, see \cite{tilloy2018exact}.

In the discrete-time formulation we divide the time $T$ in $N$ steps of duration $\dt=T/N$, $\Zcal(j)$ then reads:
\begin{equation}
    \Zcal(j) = \E{\exp\left(\sum_{k=1}^N j_k r_k\right) \Big\vert\: \rho_0},
\end{equation}
where $j_k$ is the test function value at step $k$ (at time $t=k\dt$).

To evaluate the expectation value in $\Zcal(j)$, we need to average over all possible measurement records weighted by their probability of occurrence. In the case of the jump SME, the measurement result $r_k$ at step $k$ is either $0$ or $1$, so:
\begin{align}
    \Zcal(j) = \sum_{\mathclap{r_k\in\{0, 1\}}}\> \pr{r_1, \dots, r_N \given \rho_0} \exp\left(\sum_{k=1}^N j_k r_k\right).
\end{align}

The probability of a specific measurement record is given by (\ref{eq:jump-povm-record-proba}): ${\pr{r_1, \dots, r_N \given \rho_0} = \Tr{K_{r_N}\dots K_{r_1}(\rho_0)}}$. Now we split the exponential, reorder and regroup the terms to get the final result:
\begin{align}
    \Zcal(j) &= \sum_{\mathclap{r_k\in\{0, 1\}}}\> \Tr{K_{r_N}\dots K_{r_1}(\rho_0)}\prod_{k=1}^N\exp(j_kr_k) \\
    \begin{split}
        &= \mathrm{Tr}\Bigg[\left(\sum_{r_N\in\{0, 1\}}\!\!\! K_{r_N} e^{j_N r_N}\right) \dots\\
        &\qquad\qquad\qquad \dots\left(\sum_{r_1\in\{0, 1\}}\!\!\! K_{r_1} e^{j_1 r_1}\right) (\rho_0)\Bigg]
    \end{split}\\
    &= \Tr{(K_0 + K_1 e^{j_N}) \dots (K_0 + K_1 e^{j_1}) (\rho_0)} \\
    &= \Tr{\Phi_{j_N} \dots \Phi_{j_1} (\rho_0)},
\end{align}
with $\Phi_{j_k}$ a linear map defined by:
\begin{align}
    &\Phi_{j_k}(\rho) = (K_0 + K_1 e^{j_k})(\rho)\notag\\
    &= \rho + \left[\Lcal(\rho) + (e^{j_k} - 1)\left(\theta\rho + \eta\Lt(\rho)\right)\right] \dt,\label{eq:Phi_j}
\end{align}
where we defined the superoperator ${\Lt(\rho)=L\rho\Ld}$.

By taking the limit of infinitesimally small time step, we have:
\begin{align}
    \Zcal(j) &= \Tr{\Tcal\exp\left(\int_0^T\Ls_{j_u} \dd u\right) (\rho_0)},
\end{align}
with $\Ls_{j_t}$ identified from~(\ref{eq:Phi_j}) as the generator of the evolution:
\begin{equation}\label{eq:ls-jump}
    \Ls_{j_t} = \Lcal + (e^{j_t} - 1)\left(\theta\Id + \eta \Lt\right).
\end{equation}

The calculations are very similar in the case of the diffusive SME: replacing sums by integrals over the Gaussian measure~(\ref{eq:gaussian-measure}) and using the partial Kraus map defined by~(\ref{eq:diff-povm-K}) and~(\ref{eq:diff-povm-kraus}), we obtain the same formula for $\Zcal(j)$ where the generator of the evolution $\Ls_{j_t}$ is now:
\begin{equation}\label{eq:ls-diff}
    \Ls_{j_t} = \Lcal + \sqrt{\eta}{j_t}\Lp + \frac{{j_t}^2}{2}\Id,
\end{equation}
where we defined the superoperator ${\Lp(\rho)=L\rho+\rho\Ld}$.

Note that the expression of $\mathcal{Z}(j)$ depends only on the initial state and on the SME describing the system.

\subsection{Sharp signal correlation functions}
In this subsection, we explain how to calculate the correlation functions of the sharp signal using the analytical formula for $\Zcal(j)$. The calculations for filtered signals are discussed in section~\ref{sec:4}.

We assume that the Lindbladian does not depend on time to simplify the expressions (the extension to time-dependent Lindbladian is straightforward). In the following calculations we also use the trace-preserving property of the Lindbladian evolution: ${\Tr{e^{t\Lcal}(\rho)}=\Tr{\rho}}=1$.

\textbf{Signal mean} -- The one-point correlation function for the jump SME reads:
\begin{align}
    C_{t} &= \frac{\delta}{\delta j_t} \Zcal(j) \Big\vert_{j=0}\notag\\
    \begin{split}
        &= \mathrm{Tr}\Bigg[\Tcal\exp\left(\int_t^T \Ls_{j_u} \dd u\right) e^{j_t}(\theta\Id+\eta\Lt)\notag\\
        &\qquad\qquad\qquad\qquad\;\; \Tcal\exp\left(\int_0^t \Ls_{j_u} \dd u\right)(\rho_0)\Bigg]\Bigg\vert_{j=0}
    \end{split}\\
    &= \theta + \eta\Tr{\Lt e^{t\Lcal}(\rho_0)}.
\end{align}

And for the diffusive SME:
\begin{align}
    C_{t} = \sqrt{\eta}\Tr{\Lp e^{t\Lcal}(\rho_0)}.
\end{align}

\textbf{Signal autocorrelation} -- For $t_1 < t_2$, the two-point correlation function for the jump SME reads:
\begin{align}
    C_{t_1, t_2} =&~ \frac{\delta}{\delta j_{t_1}}\frac{\delta}{\delta j_{t_2}} \Zcal(j) \Big\vert_{j=0} \notag \\
    =&~ \Tr{(\theta\Id+\eta\Lt) e^{(t_2-t_1)\Lcal} (\theta\Id+\eta\Lt) e^{t_1\Lcal} (\rho_0)}\notag \\
    =&~ \theta^2 + \eta^2 \Tr{\Lt e^{(t_2-t_1)\Lcal} \Lt e^{t_1\Lcal} (\rho_0)}\label{eq:jump-2pt}\\
     &+ \theta\eta \left(\Tr{\Lt e^{t_1\Lcal} (\rho_0)} + \Tr{\Lt e^{t_2\Lcal} (\rho_0)}\right).\notag
\end{align}

And for the diffusive SME:
\begin{align}
    C_{t_1, t_2} = \eta \Tr{\Lp e^{(t_2-t_1)\Lcal} \Lp e^{t_1\Lcal} (\rho_0)}.
    \label{eq:diff-2pt}
\end{align}

\textbf{Multipoint correlation function} -- More generally for distinct times ${t_1<\dots<t_n}$, the $n$-point correlation function for the jump SME reads:
\begin{align} \label{eq:jump-npt}
    \begin{split}
        C_{t_1, \dots, t_n} &= \mathrm{Tr}\Big[(\theta\Id+\eta\Lt) e^{(t_n-t_{n-1})\Lcal} \dots\\
        &\qquad\qquad\qquad \dots(\theta\Id+\eta\Lt) e^{t_1\Lcal} (\rho_0)\Big].
    \end{split}
\end{align}
And for the diffusive SME:
\begin{equation} \label{eq:diff-npt}
    C_{t_1, \dots, t_n} = \eta^{n/2} \Tr{\Lp e^{(t_n-t_{n-1})\Lcal} \dots \Lp e^{t_1\Lcal} (\rho_0)}.
\end{equation}

The exact result is thus obtained by inserting specific superoperators at the correlation times ($\theta I +\eta\Lt$ for the jump SME and $\sqrt{\eta}\Lp$ for the diffusive SME), and evolving the system with the ensemble-averaged evolution in-between.

\subsection{Generalisation to mixed jump-diffusive SME and multiple detectors}
We generalise the analytical formula of $\Zcal(j)$ to the case of mixed jump-diffusive SME with multiple detectors, when the quantum system is continuously measured by $n_\mu$ detectors with discrete-valued measurement results, resulting in a jump-type evolution, and by $n_\nu$ detector with continuous-valued measurement results, resulting in a diffusive-type evolution.

The general jump-diffusive SME with multiple detectors reads:
\begin{align}
    \dd\rho_t &= -i[H,\rho_t]\dt + \sum_{\mu} \mathcal{D}[V_\mu](\rho_t)\dt + \sum_\nu \mathcal{D}[L_\nu](\rho_t)\dt \notag\\
    &+ \sum_\mu\mathcal{G}[V_\mu](\rho_t)\left(\dd N_{\mu,t} - \left(\theta_\mu + \eta_\mu\Tr{V_\mu\rho_t V_\mu^\dag}\dt\right)\right)\notag\\
    &+ \sum_\nu\sqrt{\eta_\nu}\mathcal{M}[L_\nu](\rho_t)\dd W_{\nu,t},
\end{align}
where $\dd N_{\mu,t}$ are independent stochastic increments and $\dd W_{\nu,t}$ are independent Wiener processes. The observer has access to $n_\mu + n_\nu$ signals:
\begin{align}
    \Bigg\{I_{\mu, t}&=\frac{\dd Y_{\mu,t}}{\dt},J_{\nu, t}=\frac{\dd Z_{\nu,t}}{\dt}\Bigg\}_{\mu\in\llbracket 1,n_\mu\rrbracket, \nu\in\llbracket 1,n_\nu\rrbracket},\\
\shortintertext{with}
    \dd Y_{\mu,t} &= \dd N_{\mu,t},\\
    \dd Z_{\nu,t} &= \sqrt{\eta_\nu} \Tr{(L_\nu+L_\nu^\dag)\rho_t}\dt + \dd W_{\nu, t}.
\end{align}

The generating functional $\Zcal(j)$ has the same expression, where $j$ is now the set of test functions each associated with a detector: $j=\{j_\mu, j_\nu\}_{\mu\in\llbracket 1,n_\mu\rrbracket, \nu\in\llbracket 1,n_\nu\rrbracket}$. The generator of the evolution $\Ls_{j_t}$ reads:
\begin{align}
    \begin{split}
        \Ls_{j_t} =&~ \Lcal + \sum_\mu (e^{j_{\mu,t}} - 1)\left(\theta_\mu\Id + \eta_\mu V_{\mu, \times}\right) \\
        &+ \sum_\nu\left(\sqrt{\eta_\nu}{j_{\nu, t}}L_{\nu, +} + \frac{{j_{\nu, t}}^2}{2}\Id\right).
    \end{split}
\end{align}
For example, the two-point correlation function for ${t_1<t_2}$ between the jump-type detector indexed $\mu$ and the diffusive-type detector indexed $\nu$ reads:
\begin{align}
    \begin{split}
        \E{I_{\mu, t_1}J_{\nu, t_2}} &= \mathrm{Tr}\big[\sqrt{\eta_\nu}L_{\nu, +} e^{(t_2-t_1)\Lcal}\\
        &\qquad\qquad (\theta_\mu\Id+\eta_\mu V_{\mu,\times}) e^{t_1\Lcal} (\rho_0)\big].
    \end{split}
\end{align}

\section{Practical computation for realistic data in the diffusive case}\label{sec:4}
In this section we detail a novel numerical method for practically computing analytical correlation functions on experimental data. We use a simple but easily generalisable example to explain the methodology.

In a common experimental setup, the detector consists of a chain of finite bandwidth amplifiers concluded by an analogue-to-digital converter (ADC), which converts the analogue signal into a discrete signal. This amplified and digitised output signal is usually integrated against a rectangular window of duration $\Delta t$ much longer than the inverse of the ADC sampling rate. Thus from a practical point of view, the discrete-time signal $I_k$ available to an experimenter is simply the integral of the continuous-time signal $I_t$ against a rectangular window of duration $\Delta t$ (a time bin):
\begin{equation}\label{eq:binning}
    I_k = \int_{k\Delta t}^{(k+1)\Delta t} \dYt.
\end{equation}
Let us illustrate how to evaluate the two-point correlation function of this integrated signal in a slightly more general setting, when the integration windows partially overlap. This example illustrates the importance of not missing the equal time contributions when evaluating correlation functions of the filtered signal. We consider the signals $I_1$ integrated on some time interval $\Omega_1$ and $I_2$ integrated on some time interval $\Omega_2$ (see figure \ref{fig:overlap}):
\begin{align}
    I_m = \int_{t\in\Omega_m} \dYt = \int \1_{\Omega_m}(t) \dYt,
\end{align}
where $\1_\Omega$ is the rectangular window defined by $\1_\Omega(t)=1$ if $t\in\Omega$ and $\1_\Omega(t)=0$ otherwise.

\begin{center}
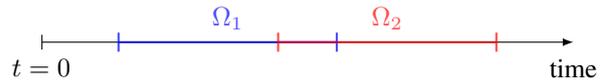

    \begin{tikzpicture}
        \draw[|-latex] node[below=3pt]{$t=0$} (0,0) -- (7,0) node[below=3pt]{time};
        \draw[thick, |-|, color=blue, opacity=0.7] (1,0) -- (3.9,0) node[above=1pt, midway]{$\Omega_1$};
        \draw[thick, |-|, color=red, opacity=0.7]  (3.1,0) -- (6,0) node[above=1pt, midway]{$\Omega_2$};
    \end{tikzpicture}
    \captionof{figure}{Overlapping integration windows}
    \label{fig:overlap}
\end{center}

We consider a system whose evolution is described by a diffusive SME, and for simplicity we assume that the Lindbladian does not depend on time.

It is tempting to evaluate the two-point correlation function naively by taking partial derivatives of the generating functional explicitly as in (\ref{eq:filtered-correlation}):
\begin{align*}
    C&_{I_1,I_2} = \pa1 \pa2 \Zcal(\alpha_1\1_{\Omega_1} + \alpha_2\1_{\Omega_2}) \Big\vert_{\alpha_1,\alpha_2=0}\\
    =&~ \int_{\Omega_1\cap\:\Omega_2} \dt\\
    &+ \eta \iint_{\Omega_1, \Omega_2, t_1\leq t_2}\dt_1 \dt_2\Tr{\Lp e^{(t_2-t_1)\Lcal}\Lp e^{t_1\Lcal} (\rho_0)} \\
    &+ \eta \iint_{\Omega_1, \Omega_2, t_1> t_2}\dt_1 \dt_2\Tr{\Lp e^{(t_1-t_2)\Lcal}\Lp e^{t_2\Lcal} (\rho_0)},
\end{align*}
where we use the abbreviated notation $\partial_{\alpha} = \frac{\partial}{\partial \alpha}$ for the partial derivative. Note the overlapping term for ${t\in\Omega_1\cap\:\Omega_2}$ coming from the equal time contributions of the sharp signal. One could then evaluate this expression numerically by i) discretising the double integrals ii) evaluating the trace integrand at each quadrature point (\textit{e.g.} by diagonalising $\mathcal{L}$ and evaluating the exponentials exactly). This is prohibitively expensive for large Hilbert space dimensions and correlation functions involving more than two points.

We propose a faster way to compute such correlation functions, which requires no discretisation when the filter is a simple binning as we assume here. We go back to the generating functional and pull the derivatives inside the trace:
\begin{align}
    C_{I_1,I_2} &= \pa1\pa2 \Tr{\Tcal\exp\left(\int_0^T\Ls_{j_u} \dd u\right) (\rho_0)}\Bigg\vert_{\alpha_1,\alpha_2=0} \notag\\
    &= \Tr{\pa1\pa2\rho_T^{j}\Big\vert_{\alpha_1,\alpha_2=0}} = \Tr{\rho_T^{(1,2)}},
\end{align}
where $j=\alpha_1 \1_{\Omega_1} + \alpha_2 \1_{\Omega_2}$ and $\rho_t^j$ is the solution to the ordinary differential equation (ODE) $\dd \rho_t^j / \dt = \Ls_{j_t}(\rho^j_t)$.
To compute the derivatives of $\rho_T^j$ with respect to $\alpha_1,\alpha_2$, we simply (forward) differentiate the ODE:
\begin{equation}
    \dodt\rho_t^{(1,2)} = \pa1\pa2\Big(\Ls_{j_t}(\rho_{t}^j)\Big)\Big\vert_{\alpha_1,\alpha_2=0}.
\end{equation}
Introducing the partial derivatives
\begin{equation}
    \rho_t^{(1)}=\pa1 \rho^j_t\Big\vert_{\alpha_1,\alpha_2=0} ~~\text{and}~~ \rho_t^{(2)}=\pa2 \rho^j_t\Big\vert_{\alpha_1,\alpha_2=0} \, ,\notag
\end{equation}
and using the explicit expression of $\Ls_{j_t}$ from (\ref{eq:ls-diff}), we obtain the system of coupled linear ODEs describing the evolution of four fictitious states $\rho_t$, $\rho_t^{(1)}$, $\rho_t^{(2)}$ and $\rho_t^{(1,2)}$:
\begingroup
\allowdisplaybreaks
\begin{align}
    \dodt\rho_t =&~ \Ls_{j_t}(\rho_{t}^j)\Big\vert_{\alpha_1,\alpha_2=0} = \Lcal(\rho_t), \label{eq:rho0}\\
    \dodt\rho_t^{(1)} =&~ \pa1\Big(\Ls_{j_t}(\rho_{t}^j)\Big)\Big\vert_{\alpha_1,\alpha_2=0} \notag \\
    =&~ \Lcal(\rho_t^{(1)}) + \1_{\Omega_1}(t)\sqrt{\eta}\Lp(\rho_t), \\
    \dodt\rho_t^{(2)} =&~ \pa2\Big(\Ls_{j_t}(\rho_{t}^j)\Big)\Big\vert_{\alpha_1,\alpha_2=0} \notag \\
    =&~\Lcal(\rho_t^{(2)}) + \1_{\Omega_2}(t)\sqrt{\eta}\Lp(\rho_t), \\
    \dodt\rho_t^{(1,2)} =&~ \pa1\pa2\Big(\Ls_{j_t}(\rho_{t}^j)\Big)\Big\vert_{\alpha_1,\alpha_2=0} \notag \\
    =&~\Lcal(\rho_t^{(1, 2)}) + \1_{\Omega_1}(t)\sqrt{\eta}\Lp(\rho_t^{(2)})\label{eq:rho12} \\
    &+ \1_{\Omega_2}(t)\sqrt{\eta}\Lp(\rho_t^{(1)}) + \1_{\Omega_1 \cap\:\Omega_2}(t)\rho_t.\notag
\end{align}
\endgroup
To obtain $\rho_T^{(1,2)}$, we solve this system linear ODE (which we see simply as a larger linear ODE) with initial condition $\rho_{t=0} = \rho_0$ and $\rho_{t=0}^{(1)}=\rho_{t=0}^{(2)}=\rho^{(1,2)}_{t=0} = 0$ from time $0$ to $T$.

This is particularly economical numerically, because the generator of the linear ODE is piecewise constant. We may thus obtain the final state by four successive exponentiation of the generator, corresponding to the evolution before $\Omega_1$, on $\Omega_1\backslash \Omega_2$, on $\Omega_1 \cap\: \Omega_2$ and finally on $\Omega_2 \backslash \Omega_1$ (the evolution for times after $\Omega_2$ is trace-preserving and thus does not need to be computed).

For large Hilbert space dimensions, one does not need to compute the exponential explicitly, but simply its action on the initial state. This can be done efficiently using Krylov subspace methods. We propose an example implementation in an elementary Julia script available at \cite{guilmincorrelations} using the QuantumOptics.jl library \cite{kramer2018quantumoptics} and the KrylovKit.jl library \cite{krylovkit}.

For more general filters, when the generator is not piecewise constant, our method remains practical and one may simply solve the time-dependent ODEs \eqref{eq:rho0} to \eqref{eq:rho12} with a high-order Runge-Kutta discretisation.

\section{Application}\label{sec:5}
The main practical application of these calculations is to infer, from $n$-point correlations given by experimental measurement data, some key parameters appearing in the modelling SME (\textit{e.g.} typical transition frequency in the Hamiltonian $H$, detection efficiency $\eta$ or dark count rate $\theta$).

Our formulation enables efficient fitting and optimisation algorithms, based for example on gradient computations via adjoint methods. Moreover, imperfections of the detection chain can be directly included in the model via the filter function $f$. In the end, this makes the estimation of parameters from realistic experimental correlation functions practical, even for Hilbert space dimensions $\sim 100$ typically arising in bosonic problems.

\section*{Acknowledgments}
This project has received funding from the European Research Council (ERC) under the European Union’s Horizon 2020 research and innovation program (grant agreement No. 884762).

\printbibliography

\end{multicols}
\end{document}